\documentclass[prd,superscriptaddress,amsfonts,amssymb,amsmath,showpacs,twocolumn]{revtex4-2}
\usepackage{bm}
\usepackage{amsfonts}
\usepackage{latexsym}
\usepackage[latin1]{inputenc}
\usepackage{graphicx}
\usepackage{amsmath,eqnarray}
\usepackage{palatino}
\usepackage{mathpazo}
\usepackage{textcomp}
\usepackage{float}
\usepackage{booktabs}
\usepackage{dcolumn}
\usepackage{hyperref}
\usepackage{mathtools}
\hypersetup{colorlinks,citecolor=red}
\hypersetup{colorlinks=true,linkcolor=blue,filecolor=blue,    urlcolor=magenta}
\usepackage{amsmath}
\usepackage{xcolor}
\usepackage{orcidlink}
\usepackage[caption=false]{subfig}
\usepackage{commath}
\def\jnl@style{\it}
\def\aaref@jnl#1{{\jnl@style#1}}

\def\aaref@jnl#1{{\jnl@style#1}}

\def\aj{\aaref@jnl{AJ}}                   % Astronomical Journal
\def\apj{\aaref@jnl{ApJ}}                 % Astrophysical Journal
\def\apjl{\aaref@jnl{ApJ}}                % Astrophysical Journal, Letters
\def\apjs{\aaref@jnl{ApJS}}               % Astrophysical Journal, Supplement
\def\apss{\aaref@jnl{Ap\&SS}}             % Astrophysics and Space Science
\def\aap{\aaref@jnl{A\&A}}                % Astronomy and Astrophysics
\def\aapr{\aaref@jnl{A\&A~Rev.}}          % Astronomy and Astrophysics Reviews
\def\aaps{\aaref@jnl{A\&AS}}              % Astronomy and Astrophysics, Supplement
\def\mnras{\aaref@jnl{Mon.~Not.~Roy.~Astron.~Soc.}}             % Monthly Notices of the RAS
\def\prd{\aaref@jnl{Phys.~Rev.~D}}        % Physical Review D
\def\prc{\aaref@jnl{Phys.~Rev.~C}}  % Physical Review C
\def\prl{\aaref@jnl{Phys.~Rev.~Lett.}}    % Physical Review Letters
\def\qjras{\aaref@jnl{QJRAS}}             % Quarterly Journal of the RAS
\def\skytel{\aaref@jnl{S\&T}}             % Sky and Telescope
\def\ssr{\aaref@jnl{Space~Sci.~Rev.}}     % Space Science Reviews
\def\zap{\aaref@jnl{ZAp}}                 % Zeitschrift fuer Astrophysik
\def\nat{\aaref@jnl{Nature}}              % Nature
\def\aplett{\aaref@jnl{Astrophys.~Lett.}} % Astrophysics Letters
\def\apspr{\aaref@jnl{Astrophys.~Space~Phys.~Res.}} % Astrophysics Space Physics Research
\def\physrep{\aaref@jnl{Phys.~Rep.}}      % Physics Reports
\def\physscr{\aaref@jnl{Phys.~Scr}}       % Physica Scripta
\def\commat{\aaref@jnl{Comm.~Math.~Phys.}}              % Communications in Mathematical Physics
\def\science{\aaref@jnl{Science}}               % Science
\def\cqg{\aaref@jnl{Classical Quant.~Grav.}}            % Classical and Quantum Gravity
\def\jpcs{\aaref@jnl{JPCS}}                                     % Journal of Physics Conference Series
\def\ijmpd{\aaref@jnl{Int.~J.~Mod.~Phys.~D}}                    % International Journal of Modern Physics D
\def\grg{\aaref@jnl{Gen.~Relat.~Gravit.}}               % General Relativity and Gravitation
\def\rpp{\aaref@jnl{Rep.~Prog.~Phys.}}          % Reports on Progress in Physics
\def\npa{\aaref@jnl{Nucl.~Phys.~A}}        % Nuclear Physics A
\def\lrr{\aaref@jnl{Living Rev.~Rel.}}                   % Living reviews in relativity
\def\jcap{\aaref@jnl{J.~Cosmology Astropart.~Phys.}}    % Journal of cosmology and astroparticle physics
\def\rmp{\aaref@jnl{Rev.~Mod.~Phys.}}   %Reviews of modern physics
\def\epjc{\aaref@jnl{Eur.~Phys.~J.~C}}

\allowdisplaybreaks[1]
\begin{document}

\color{black}  
\title{Properties of wormhole model in de Rham-Gabadadze-Tolley like massive gravity with specific matter density}

\author{Piyali Bhar \orcidlink{0000-0001-9747-1009} }
\email{piyalibhar90@gmail.com}
\affiliation{Department of
Mathematics, Government General Degree College Singur, Hooghly, West Bengal 712409,
India}

\begin{abstract}
In the conventional method of studying wormhole (WH) geometry, traversability requires the presence of exotic matter, which also provides negative gravity effects to keep the wormhole throat open. In dRGT massive gravity theory, we produce two types of WH solutions in our present paper. Selecting a static and spherically symmetric metric for the background geometry, we obtain the field equations for exact WH solutions. We derive the WH geometry completely for the two different choices of redshift functions. All the energy conditions including the NEC are violated by the obtained WH solutions. Various plots are used to illustrate the behavior of the wormhole for a suitable range of $m^2c_1$, where $m$ is the graviton mass. It is observed that the photon deflection angle becomes negative for all values of $m^2c_1$ as a result of the repulsive action of gravity. It is also studied that the repulsive impact of massive gravitons pushes the spacetime geometry so strongly that the asymptotic flatness is affected. The Volume Integral Quantifier (VIQ) has also been computed to determine the amounts of matter that violate the null energy condition. The complexity factor of the proposed model is also discussed.


\end{abstract}

\maketitle
%\end{CJK*}

\section{Introduction}
Wormholes are tunnel-shaped geometric structures that connect two different universes or two different places inside the same universe. The models of wormholes were initially proposed by Flamm \cite{flamm} when he established the Schwarzschild solution for isometric embedding. Following Flamm's \cite{flamm} research, Einstein and Rosen \cite{Einstein:1935tc} developed a geometric structure that is now known as the Einstein-Rosen bridge. It is interesting to note that the term ``wormhole" was coined by Wheeler \cite{Wheeler:1955zz} for the first time in 1957 and then by Misner and Wheeler \cite{Misner:1957mt}. Morris and Thorne \cite{Morris:1988cz} initially proposed the concept of a traversable wormhole. They studied spherically symmetric static objects using General Relativity (GR) and demonstrated the need to address energy conditions. Physical properties of this exotic matter that violate energy conditions, such as a particle with negative mass, would be incompatible with the standard law of physics. Following that, the theory by Morris and associates \cite{Morris:1988cz,Morris:1988tu} suggested that a wormhole might exist and eventually evolve into a time machine capable of overcoming causality. The minimization of exotic matter is a major difficulty in wormhole physics. Various strategies have been developed in the literature to reduce the use of exotic matter, such as the ``cut and paste"  method \cite{Visser:1989kh,Visser:1989kg}. However, this technique is limited to wormhole throats. Additionally, an efficient measure known as the ``volume integral quantifier" (VIQ) has been devised by Visser et al. \cite{Visser:2003yf} to determine the amount of exotic matter needed for a traversable wormhole.\par

Since the traversable wormhole model was proposed, it has become a topic of interest if wormholes can be generated from regular matter. It has been demonstrated that modified theories of gravity might be able to reduce or perhaps eliminate the requirement for exotic matter significantly. The additional degrees of freedom in these theories lead to the appearance of new terms in their field equations that permit the formation of wormholes while maintaining energy requirements. Developing traversable wormholes in the $f(R)$ gravity, where $R$ is the Ricci scalar, Mazharimousavi and Halilsoy \cite{Mazharimousavi:2016npo} demonstrated that the wormholes respect at least the weak energy requirement. Additionally, DeBenedictis and Horvat \cite{DeBenedictis:2012qz} have created $f(R)$ wormholes that adhere to energy constraints. In a recent study \cite{Roy:2022kid}, the wormhole model in modified gravity and Einstein gravity theories were discussed. It was found that in modified gravity, it is possible to have a wormhole with regular matter that satisfies all energy conditions. Wormholes with non-exotic matter were discovered through several ground-based investigations \cite{Fukutaka:1989zb,Hochberg:1990is,Kanti:2011yv,Harko:2013yb,Sengupta:2021wvi}. Notably, wormholes in modified gravity present a novel development in the field of applications of gravitational theories. Modified gravity has become a popular substitute for dark energy and dark matter in recent years \cite{Bertschinger:2008zb,Lue:2003ky,Langlois:2018dxi,Sanders:2006sz,Cembranos:2015svp,Rinaldi:2016oqp}. Basically, modified theories of gravity (MTGs) are geometric extensions of Einstein's General Relativity, where cosmic acceleration can be obtained through the modification of the Einstein-Hilbert action integral. As MTG helps us to understand cosmic expansion and other related concepts, it would be interesting to experiment with the capacity of these theories to study astrophysical objects like wormholes, compact stars, etc. In particular, early and late time accelerated expansion of the Universe has been described by MTG. Recently, the authors of refs. \cite{Tsukamoto:2012xs,Shaikh:2018kfv,Gyulchev:2018fmd} have published some intriguing findings on the shadows of wormholes. Additionally, one might go through some amazing papers on wormhole geometries in other MTG, including those in the fields of $f(R)$ gravity \cite{Lobo:2009ip,Bronnikov:2010tt,Sharif:2018jdj,Shamir:2021fkg}, $f(R,\,T)$ gravity \cite{Sahoo:2019aqz,Sharif:2019nuv,Bhar:2021lat,Godani:2021gri}, $f(T)$ gravity \cite{Sharif:2014rda,Sharif:2013exa,Jawad:2022aky,Ditta:2021uoe}, and other modified theories gravity \cite{Kavya:2024kdi,Maurya:2024jos,Kiroriwal:2024moj,Tayde:2024shx,Khatri:2024sdi,Mustafa:2024iwu}.\par

Pauli and Fierz \cite{Fierz:1939ix} were the creators of one such theory, known as massive gravity. Incorporating a spin-2 mass field into this theory, particularly the assignment of mass to the graviton, became an important point in explaining the recent accelerating expansion at large distances without requiring dark energy. It was discovered that their theory suffered from a discontinuity in its predictions-the so-called van Dam-Veltman-Zakharov (vDVZ) discontinuity-which was identified by van Dam, Veltman, and Zakharov \cite{VanNieuwenhuizen:1973fi,vanDam:1970vg,Zakharov:1970cc}. Further investigation on the nonlinear generalization of Fierz-Pauli's massive gravity was driven by this discontinuity issue. While looking for such generalizations, Vainshtein discovered that the prediction made by linearized theory is not reliable inside a characteristic radius known as the ``Vainshtein" radius and proposed the idea of the nonlinear massive gravity mechanism which can be utilized to restore predictions of general relativity \cite{Vainshtein:1972sx}.
Boulware and Deser discovered that these nonlinear generalizations typically result in an equation of motion with a higher derivative term, which causes ghost instability in the theory that is thereafter referred to as a Boulware-Deser (BD) ghost \cite{Boulware:1972yco}. However, by first introducing St\"{u}ckelberg fields, these issues that occurred during the formation of the massive gravity have been tackled in the last decade \cite{Arkani-Hamed:2002bjr}. This allows for a class of potential energies that rely on an internal Minkowski metric and the gravitational metric.
Moreover, the set of permitted mass terms was restricted and perturbatively supplied by deRham, Gabbadze, and Tolley (dRGT) to prevent the reappearance of the ghost in massive gravity \cite{deRham:2011by,deRham:2010ik,deRham:2014zqa}. After adding these terms together, they discovered three possible combinations of the mass terms: quadratic, cubic, and quadratic. The dRGT massive gravity is appropriately developed to eliminate the higher derivative term from the equations of motion and eliminate the ghost field. The de Rham-Gabadadze-Tolley (dRGT) massive theory is one of the most interesting options among the modified theories of gravity when examining the universe at the cosmic scale. Ghosh et al. \cite{Ghosh:2015cva} provided a precise spherical black hole solution in dRGT massive gravity. For a generic choice of the parameters, the authors also covered the thermodynamics and phase structure of the black hole. In dRGT massive gravity, Chabab et al. \cite{Chabab:2019mlu} investigated the thermodynamics and geothermodynamics of spherical black hole solutions in a newly extended phase space. The authors of this study use a thermodynamical analysis to show that black holes in dRGT massive gravity exhibit a critical behavior with two distinct critical points via canonical and grand canonical ensembles. The model of compact star and dark energy star in dRGT massive gravity can be found in \cite{Tudeshki:2023ias,Li:2023rkr}. In addition, the spherically symmetric solutions for dRGT gravity were discussed in \cite{Nieuwenhuizen:2011sq,Brito:2013xaa}, and in ref. \cite{Berezhiani:2011mt}, the equivalent charged black hole solution was discovered.\\

In General Relativity, to obtain the traversable wormhole model, the null energy criterion must be violated. The massive graviton density can be negative in a certain region where the radial pressure is positive. Therefore, the massive graviton energy-momentum tensor in the massive gravity theory readily exhibits the violation of energy condition \cite{Sushkov:2015fma}. Since wormholes can naturally arise from massive gravity, the description of this specific object has major importance on its own. Here we are describing some earlier work done on wormhole solutions in dRGT massive gravity with certain restrictions. In $f(R)$ massive gravity, Tangphati et al. \cite{Tangphati:2020mir} studied the traversable wormhole solution and discussed the asymptotic geometry by considering an exponential shape function. However, they could not produce a geometry that takes the influence of massive gravitons into account because of their limited options for shape functions. In addition, Kamma et al. \cite{Kamma:2021wam} reported the wormhole solution in massive gravity theory and got the shape function, as well as the redshift function which contains the massive gravity parameters. The wormhole solution is analyzed by Dutta et al. \cite{Dutta:2023wfg} in both Einstein's massive gravity and the dRGT-$f(R,\,T)$ massive gravity. The shape function in both models is derived from the anisotropic pressure solution in the ultrastatic wormhole geometry, which includes the massive gravity constants $\gamma$ and $\Lambda$. Nevertheless, the terms $\gamma$ and $\Lambda$ act in a way that makes the spacetime non-asymptotically flat.  Bhar et al. \cite{Bhar:2024vov} recently proposed a model of a non-commutative wormhole in de Rham-Gabadadze-Tolley-like massive gravity and demonstrated that photons following null geodesics have a negative deflection angle when repulsive gravity is present.\par
%%%%%%%%%%%%%%%%%%
Inspired by the previous work done, in the present paper we are interested in searching for wormhole solutions in the background of dRGT massive gravity. The primary objective of our study is to determine how the characteristics of massive gravity affect wormhole spacetime. The plan of our paper is as follows: Sect.~\ref{sec2} provides an overview of the dRGT massive gravity theory and the accompanying field equations in this modified gravity. In Sect.~\ref{sec3}, we select a physically realistic matter density and two physically reasonable redshift functions that guarantee the absence of horizons and singularities in space-time. We derive the shape function of the wormhole using these particular choices of the redshift function. Sect.~\ref{sec4} verifies the standard energy conditions of the present model. In Sect.~\ref{sec5}, we have discussed the deflection angle of our proposed model. Some aspects of the wormhole solution, especially the measurement of exotic matter using the volume integral quantifier (VIQ), the complexity factor, and the TOV equations, are examined in Sects.~\ref{sec6}-\ref{sec8}. Sect.~\ref{sec9} concludes with a summary and discussion of our findings.
\section{Traversable wormhole spacetimes and field equations}\label{sec2}
A spherically symmetric, static wormhole is represented by the spacetime metric given by,
\begin{eqnarray}\label{line}
ds^2&=&-e^{2\Phi}dt^2+\frac{dr^2}{1-\frac{b(r)}{r}}+r^2(d\theta^2+\sin^2\theta d\phi^2),
\end{eqnarray}
where both $\Phi(r)$ and $b(r)$ are arbitrary functions of radial coordinate `r'. $\Phi(r)$ is referred to as the redshift function, as it is associated with the gravitational redshift. On the other hand, $b(r)$ is known as the shape function since it determines the shape of the wormhole. The range of the radial coordinate increases from a minimal value at $r_0$, which is the location of the wormhole throat, to the location of the joining of the inner spacetime with an external vacuum solution. A crucial feature of a wormhole is that it needs to meet both the $1 - \frac{b(r)}{r} >0$ and the flaring out condition at the throat, which is given by $\frac{(b-b'r)}{b^2}>0$. Moreover, wormhole solutions require that $b'(r_0)<1$ and that $b(r_0)=r_0$ at the throat. A further essential requirement is that $\Phi(r)$ must always be finite because there cannot be any horizons defined as the surfaces where $e^{2\Phi}\rightarrow~0$.\par

The action that describes the massive gravity model in this paper is as follows \cite{deRham:2010kj,Berezhiani:2011mt}:
\begin{equation}\label{eq1}
	 I=\frac{1}{16\pi }\int d^{4}x\sqrt{-g}[\mathcal{R}+m^{2}{%
\sum\limits_{i}^{4}c_{i}\mathit{\mathcal{U}_{i}(g,f)}}]+I_{matter},
\end{equation}
where $ \mathcal{R} $ is the Ricci scalar, $ m$ denotes the graviton mass, $c_{i}$ are constant coefficients and which plays the role of
free parameters of action. Also $\mathcal{U}_{i}$ are introduced as symmetric
polynomials of the eigenvalues of matrix $\mathcal{K}_{~~b}^{a}=\sqrt{g^{ac}f_{cb}}$, $ I_{matter} $ represents the matter Lagrangian, and $ g =det(g_{ab}) $, $f_{ab}$ is called the reference metric.

%In four-dimensional spacetime, the effective potential $ \mathcal{U} $ is defined as
%\begin{eqnarray}\label{potential}
%	\mathcal{U}(g,\phi^a) = \mathcal{U}_2 + \alpha_3 \mathcal{U}_3 + \alpha_4 \mathcal{U}_4 ,
%\end{eqnarray}
%here $\alpha_3$ and $\alpha_4$ are dimensionless free parameters of the theory.
Whereas, the expressions for $\mathcal{U}_{1},\,\mathcal{U}_{2},\,\mathcal{U}_{3}$ and $\mathcal{U}_{4}$ are given by
\begin{eqnarray}\label{eq3}
\mathcal{U}_1&=&[\mathcal{K}],\nonumber\\
	\mathcal{U}_2 &=& [\mathcal{K}]^2 - [\mathcal{K}^2], \nonumber \\
	\mathcal{U}_3 &=& [\mathcal{K}]^3 - 3[\mathcal{K}][\mathcal{K}^2] + 2 [\mathcal{K}^3], \nonumber \\
	\mathcal{U}_4 &=& [\mathcal{K}]^4 - 6[\mathcal{K}]^2[\mathcal{K}^2] + 8[\mathcal{K}][\mathcal{K}^3] + 3[\mathcal{K}^2]^2 -     6[\mathcal{K}^4], \nonumber \\
	%\mathcal{{K}^{\mu}}_{\nu} &=& \delta^{\mu}_{\nu} - \sqrt{g^{\mu\lambda} \partial_{\lambda}\phi^a \partial_{\nu}\phi^b \mathcal{F}_{ab} },
\end{eqnarray}
and,
%\begin{eqnarray}
 %{\cal K}^\mu_\nu =
%\delta^\mu_\nu-\sqrt{g^{\mu\sigma}f_{ab}
%\partial_\sigma\phi^a\partial_\nu\phi^b}, \label{K-tensor}
%\end{eqnarray}
Here, $[\mathcal{K}]=\mathcal{K}_{a}^{a}$ and $%
[\mathcal{K}^{n}]=(\mathcal{K}^{n})_{a}^{a}$, and the reference fiducial metric $ f_{ab} $ has the explicit form given by \cite{Vegh:2013sk,Cai:2014znn,Adams:2014vza,Xu:2015rfa},
\begin{equation}\label{eq4}
	f_{ab} =
	\begin{pmatrix}
		0 & 0 & 0 & 0 \\
		0 & 0 & 0 & 0 \\
		0 & 0 & C^2 & 0 \\
		0 & 0 & 0 & C^2 \text{sin}^2 \theta \\
	\end{pmatrix},
\end{equation}
where $C$ is a positive constant.\\
%where we choose the unitary gauge, $ \phi^a = x^\mu \delta^a_\mu $ \cite{Vegh:2013sk}, so that
%\begin{equation}\label{eq5}
%	\sqrt{g^{\mu\lambda} \partial_{\lambda}\phi^a \partial_{\nu}\phi^b f_{ab} }= \sqrt{g^{\mu \lambda} f_{\lambda \nu} }.
%\end{equation}
%Now, let us redefine the two
%parameters $\alpha_3$ and $\alpha_4$ of the graviton potential in Eq. \eqref{potential} by introducing two new parameters $\alpha$ and
%$\beta$, as follows
%\begin{eqnarray}\label{alphabeta}
 %\alpha_3 = \frac{\alpha-1}{3}~,~~\alpha_4 =
%\frac{\beta}{4}+\frac{1-\alpha}{12}.
%\end{eqnarray}
The modified Einstein field equations can be obtained by varying the action about metric $g_{\mu\nu}$ as follows:
\begin{equation}
G_{ab}+m^{2}\chi _{ab}=8\pi T_{ab},  \label{fieldeq}
\end{equation}%
where $G_{ab}$ is Einstein tensor, $\chi _{ab}$ is called massive tensor and
$T_{ab}$ is stress-energy tensor. Note; in the following, we use geometrized
units $(c=G=1)$ for calculations. The massive tensor $\chi _{ab}$ is
extracted to form
\begin{equation}
\chi _{ab}=-\sum_{i=1}^{d-2}\frac{c_{i}}{2}\left[ {\mathit{\mathcal{U}}}%
_{i}g_{ab}+\sum_{y=1}^{i}\left( -1\right) ^{y}\frac{i!}{\left( i-y\right) !}{%
\mathit{\mathcal{U}}}_{i-y}\left[ \mathcal{K}_{ab}^{y}\right] \right] ,  \label{chi}
\end{equation}%
where $d$ is related to the dimensions of spacetime. We work on
4-dimensional spacetime and so $d=4$.\\
Using the line element Eq. (\ref{line})
and reference metric Eq. (\ref{eq4}), we can determine the
tensor $\mathcal{K}_{~~b}^{a}$
\begin{equation}
\mathcal{K}_{~~b}^{a}=diag\left( 0,0,\dfrac{C}{r},\dfrac{C}{r}\right) ,
\label{k tensor}
\end{equation}%
and
\begin{eqnarray}
(\mathcal{K}^{2})_{~~b}^{a} &=&\left[
\begin{array}{cccc}
0 & 0 & 0 & 0 \\
0 & 0 & 0 & 0 \\
0 & 0 & \frac{{C}^{2}}{{r}^{2}} & 0 \\
0 & 0 & 0 & \frac{{C}^{2}}{{r}^{2}}%
\end{array}%
\right] ,  \notag \\
&&  \notag \\
(\mathcal{K}^{3})_{~~b}^{a} &=&\left[
\begin{array}{cccc}
0 & 0 & 0 & 0 \\
0 & 0 & 0 & 0 \\
0 & 0 & \frac{{C}^{3}}{{r}^{3}} & 0 \\
0 & 0 & 0 & \frac{{C}^{3}}{{r}^{3}}%
\end{array}%
\right] ,  \notag \\
&&  \notag \\
(\mathcal{K}^{4})_{~~b}^{a} &=&\left[
\begin{array}{cccc}
0 & 0 & 0 & 0 \\
0 & 0 & 0 & 0 \\
0 & 0 & \frac{{C}^{4}}{{r}^{4}} & 0 \\
0 & 0 & 0 & \frac{{C}^{4}}{{r}^{4}}%
\end{array}%
\right] ,  \label{k power}
\end{eqnarray}
and also
\begin{eqnarray}
\lbrack \mathcal{K}] &=&\frac{2C}{r},~~~\&~~~[\mathcal{K}^{2}]=\dfrac{2C^{2}}{r^{2}},  \notag \\
&&  \notag \\
\lbrack \mathcal{K}^{3}] &=&\dfrac{2C^{3}}{r^{3}},~~~\&~~~[\mathcal{K}^{4}]=\dfrac{2C^{4}}{r^{4}}%
.  \label{k trace}
\end{eqnarray}%
Since we are working on 4-dimensional spacetime, the only non-zero terms of $%
\mathcal{U}_{i}$ are $\mathcal{U}_{1}$ and $\mathcal{U}_{2}$\thinspace\ \cite{Hendi:2017fxp}. Therefore,
using Eqs. (\ref{eq3}), (\ref{k tensor}) and (\ref{k trace}), $\mathcal{U}_{i}$ are
obtained
\begin{eqnarray}
\mathcal{U}_{1} &=&\dfrac{2C}{r},~~~\&~~~\mathcal{U}_{2}=\dfrac{2C^{2}}{r^{2}},  \notag \\
\mathcal{U}_{i} &=&0,\text{ \ \ when \ \ }i>2.  \label{ui}
\end{eqnarray}%
By putting Eq. (\ref{ui}) in Eq. (\ref{chi}), the elements of the massive
tensor is determined
\begin{eqnarray}
\chi _{~~1}^{1} &=&-\dfrac{C(Cc_{2}+c_{1}r)}{r^{2}},~~~\&~~\chi _{~~3}^{3}=-%
\dfrac{c_{1}C}{2r},  \notag \\
&&  \notag \\
~\chi _{~~2}^{2} &=&-\dfrac{C(Cc_{2}+c_{1}r)}{r^{2}},~~~\&~~~\chi
_{~~4}^{4}=-\dfrac{c_{1}C}{2r}.  \label{chi1}
\end{eqnarray}
%%%%%%%%%%%%%%%%%%%%%%%%%%%%%%%%%%%%%%%%%%%
Let us assume that the pressure in our current system is anisotropic. As a result, the energy-momentum tensor $ T_{ab} $ can be expressed as follows in terms of principal pressure:
\begin{equation}\label{eq10}
	T_{ab}= (\rho+p_t)u_a u_b +p_t g_{ab} +(p_r-p_t)v_a v_b,
\end{equation}
where $ u_a $ is the timelike unit vector, $ v_a $ is a spacelike unit vector orthogonal to the timelike unit vector, such that $ u_a u^a=-1 $ and $ v_a v^a=1 $.\\
In dRGT gravity, we can express the Einstein Field Equations (EFEs) (assuming $G=c=1$) in an orthonormal reference frame using the line element (\ref{line}), which results in the following set of equations:
%\begin{widetext}
  \begin{eqnarray}
      8\pi\rho &=& \frac{b^{\prime}}{r^{2}}+\frac{m^2C(c_2C+c_1r)}{r^2}   ,  \label{l2}  \\
        8\pi p_r &=& 2\left(1-\frac{b}{r}\right)\frac{\Phi^{\prime}}{r}-\frac{b}{r^{3}}-\frac{m^2C(c_2C+c_1r)}{r^2} ,  \label{13}  \\
        8\pi p_t  &=& \left(1-\frac{b}{r}\right)\left[\Phi^{\prime\prime}+\Phi^{\prime 2}-\frac{b^{\prime}r-b}{2r(r-b)}\Phi^{\prime}\right.\nonumber\\&&\left.-\frac{b^{\prime}r-b}{2r^2(r-b)} +\frac{\Phi^{\prime}}{r}\right]-\frac{m^2c_1C}{2r}  ,  \label{14}
  \end{eqnarray}
%\end{widetext}
The conservation equation can be obtained as, $T_{a;b}^b=0$, which implies,
\begin{eqnarray}
\frac{dp_r}{dr}&=&-(p_r+\rho)\frac{d\Phi}{dr}+\frac{2}{r}(p_t-p_r).
\end{eqnarray}

\section{Two different model}\label{sec3}
\begin{figure*}
    \centering
     \includegraphics[height=6cm,width=7.5cm]{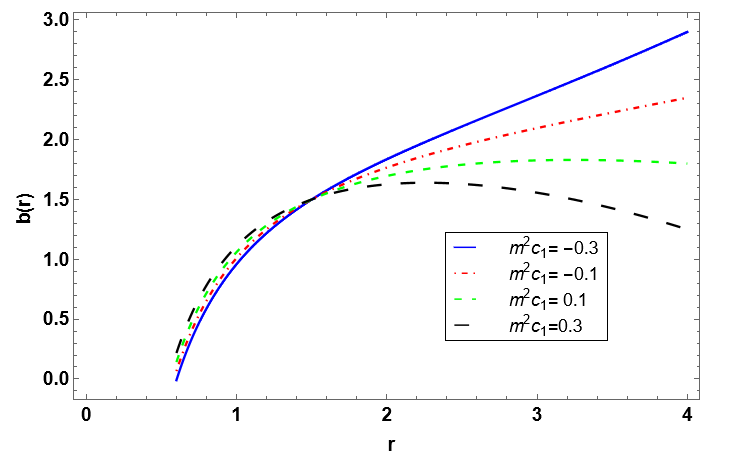}
    \includegraphics[height=6cm,width=7.5cm]{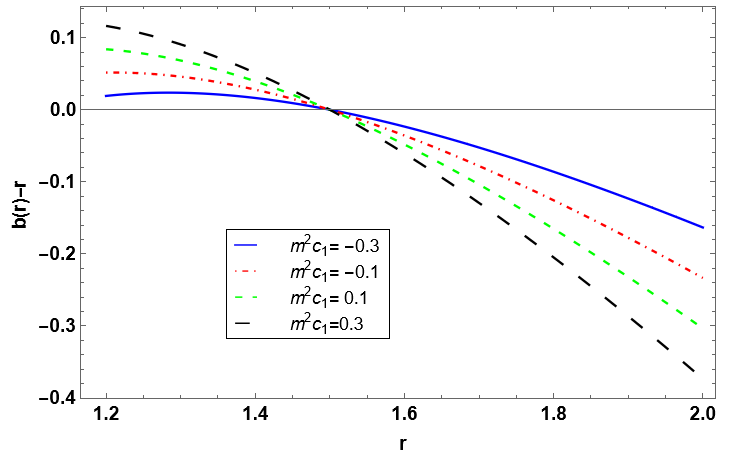}
        \caption{The shape function and position of the throat are shown against `r'. For drawing the plots we have taken the values of the parameter as $r_0 = 1.5,\,C = 0.4,\,m^2c_2 =0.02,\,\rho_c = 0.011$, and $\beta = 4$. \label{fig1}}
\end{figure*}

\begin{figure*}
    \centering
     \includegraphics[height=6cm,width=7.5cm]{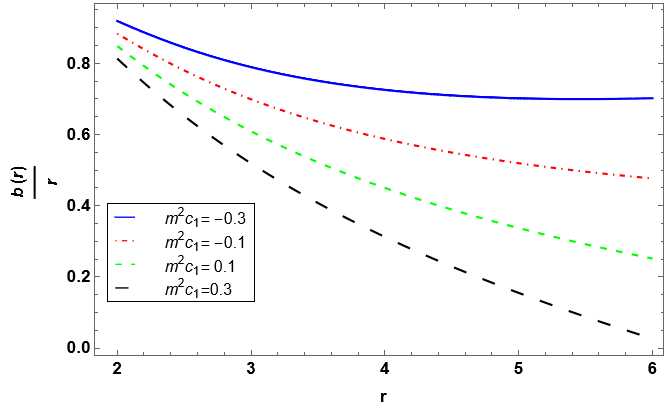}~~~
    \includegraphics[height=6cm,width=7.5cm]{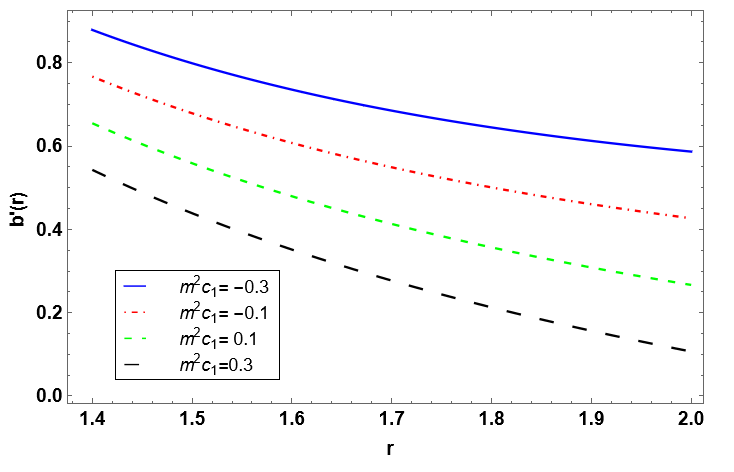}\\
 \includegraphics[height=6cm,width=7.5cm]{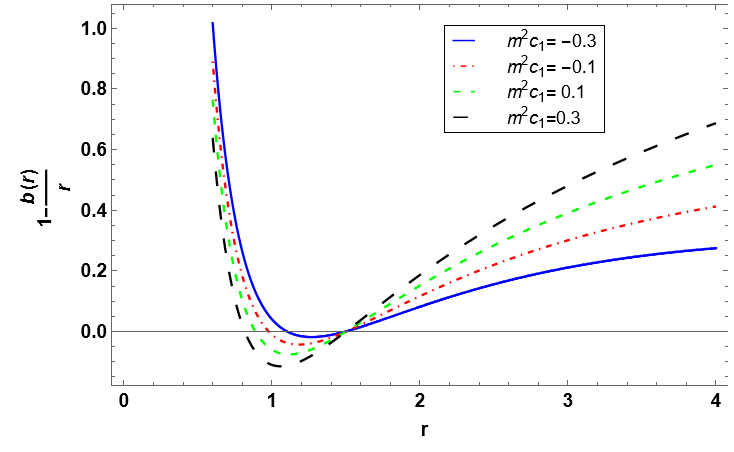}
        \caption{The asymptotically flatness condition, flaring out condition, and $1-\frac{b(r)}{r}$ are shown against r. For drawing the plots we have taken the values of the parameter as $r_0 = 1.5,\,C = 0.4,\,m^2c_2 =0.02,\,\rho_c = 0.011$, and $\beta = 4$. \label{fig2}}
\end{figure*}

In this section, we begin by determining WH solutions in the framework of massive gravity using two types of redshift functions and a scientifically realistic matter density function. We take $\Phi' = 0$ and $\Phi=\frac{A}{r}$ into consideration for the WH to be traversable. This decision ensures that the geometry of the WH is free from the horizon and that the redshift function takes finite value everywhere inside the spacetime. Subsequently, we take into account the matter density \cite{Lobo:2012qq},
\begin{eqnarray}\label{rho}
\rho=\rho_c\left(\frac{r_0}{r}\right)^{\beta},
\end{eqnarray}
with $\beta>0$. It can be observed that the energy density has a smooth profile, maximum at the throat of the wormhole, and monotonic decreasing with the increasing value of `r'. These assumptions considerably simplify the field equations and yield particularly interesting solutions, which we present below.\\
By employing the expression of $\rho$ given in equation (\ref{rho}), we solve the eqn. (\ref{l2}), which provides the expression of the shape function $b(r)$ as,
\begin{eqnarray}
b(r)=-C m^2 \left(C c_2 r + \frac{c_1 r^2}{2}\right) + \frac{8 \pi r^3 \left(\frac{r_0}{r}\right)^{\beta} \rho_c}{3 - \beta}+d,\nonumber\\
\end{eqnarray}
where $d$ is the constant of integration which can be found by using the condition $b(r_0)=r_0$ as,
\[d=r_0 + C m^2 \left(C c_2 r_0 + \frac{c_1 r_0^2}{2}\right) - \frac{8 \pi r_0^3 \rho_c}{3 - \beta}.\]
Finally, we obtain the expression for the shape function as,
%\begin{widetext}
\begin{eqnarray}\label{br}
b(r)&=&r_0 + C^2  m^2c_2 (r_0-r) + \frac{1}{2} C m^2c_1 ( r_0^2-r^2 ) \nonumber\\&&+ \frac{
 8 \pi \left[r_0^3 - r^3 \left(\frac{r_0}{r}\right)^{\beta}\right] \rho_c}{\beta-3},
\end{eqnarray}
%\end{widetext}
The radial derivative of the shape function is given as,
\[b'(r)=-C m^2 (C c_2 + c_1 r) + 8 \pi r^2 \left(\frac{r_0}{r}\right)^{\beta} \rho_c,\]
Now at the throat of the wormhole, $b'(r_0)<1$ which implies,
\[ \rho_c<\frac{1+C m^2 (C c_2 + c_1 r_0)}{8 \pi r_0^2}\]
We will next verify whether the shape function $b(r)$ (see left panel of Fig.~\ref{fig1}) meets all the necessary physical conditions to construct a structure like a wormhole. To check this, we allot some arbitrary values to the parameters (specified in the figures). When $m^2c_1$ assumes a large positive value, as seen in Fig.~\ref{fig2} (upper left panel), we can observe that $\frac{b(r)}{r}$ tends to `zero' as $r~\rightarrow~\infty$. The wormhole throat, $r_0$, is located at the point where $b(r_0) = r_0$, which is the root of the function $b(r)-r$. In Fig.~\ref{fig1} (right panel), we have shown $b(r)-r$. We may observe from this figure that $r_0 = 1.5$ regardless of $m^2c_1$ values. Additionally, Fig.~\ref{fig2} (bottom panel) shows that for $r > r_0$, $1-\frac{b(r)}{r}>0$. Furthermore, Fig.~\ref{fig2} (upper right panel) shows that $b'(r)<1$ near the throat of the wormhole $r_0 = 1.5$. As a result, the flare-out requirement is met as well.\par
\begin{figure*}
    \centering
     \includegraphics[height=6cm,width=7.5cm]{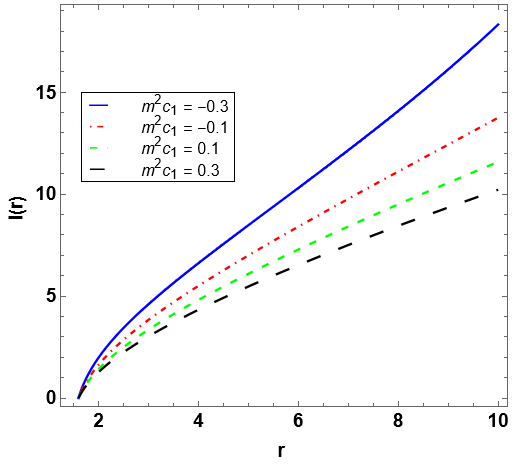}
    \includegraphics[height=6cm,width=7.5cm]{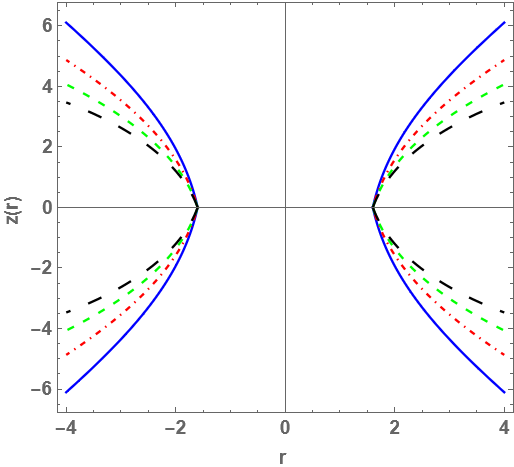}
        \caption{The proper radial distance and embedding diagram of the wormhole are shown in the figure. For drawing the plots we have taken the values of the parameter as $r_0 = 1.5,\,C = 0.4,\,m^2c_2 =0.02,\,\rho_c = 0.011$, and $\beta = 4$.\label{zr}}
\end{figure*}

\begin{figure}
    \centering
     \includegraphics[height=8.2cm,width=8.6cm]{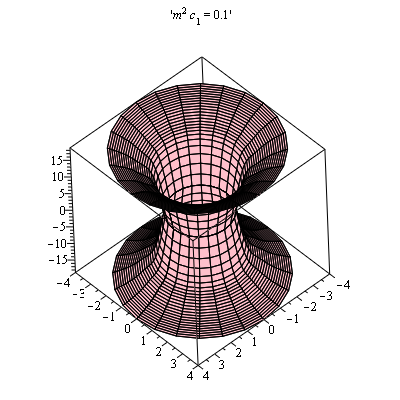}
        \caption{Full visualization of the wormhole. For drawing the plots we have taken the values of the parameter as $A=0.3,\,r_0 = 1.5,\,C = 0.4,\,m^2c_2 =0.02,\,\rho_c = 0.011$, and $\beta = 4$.\label{full}}
\end{figure}

%%%%%%%%%%%%% embedding surface
We utilize geometrical embedding diagrams to provide better visualization. The expression of the shape function $b(r)$ plays a crucial role in determining the features of these diagrams. We focus our analysis on the equatorial slice that is defined by $\theta=\pi/2$ and a fixed time coordinate, $t=$constant.
In these particular circumstances, the metric obtained from (\ref{line}) takes on the following form:
\begin{eqnarray}\label{sur1}
ds^2&=&\frac{dr^2}{1-\frac{b(r)}{r}}+r^2 d\phi^2,
\end{eqnarray}
This slice can then be inserted into its hypersurface using cylindrical coordinates ($r,\,\phi,\,z$) to get the metric:
\begin{eqnarray}\label{sur2}
ds^2&=&dz^2+dr^2+r^2 d\phi^2,
\end{eqnarray}
 %By consulting [M.S. Morris, K.S. Thorne, Amer. J. Phys. 56 (1988) 395], the expression for the embedding surface $z(r)$ can be obtained as,
Now equating eqns. (\ref{sur1}) and (\ref{sur2}), we finally obtain the following equation:
\begin{eqnarray} \label{z}
z(r)=\pm\int_{r_0^{+}}^{r}\left(\frac{r}{b(r)}-1\right)^{-\frac{1}{2}}dr,
\end{eqnarray}
Using Eq. (\ref{z}), we depict the embedding diagram in Fig.~\ref{zr} (right panel) for different values of $m^2c_1$ and through a $2\pi$ rotation around the z-axis for $m^2c_1=0.1$ is shown in Fig.~\ref{full}.\par
One can obtain the proper radial distance $l(r)$ of wormhole as follows:
\begin{eqnarray}
l(r)&=&\int_{r_0^{+}}^r\frac{dr}{\sqrt{1-\frac{b(r)}{r}}},
\end{eqnarray}
The aforementioned integral cannot be calculated analytically. In Fig.~\ref{zr} (left panel), we presented a graphical representation of the proper radial distance using a numerical integral. It is noteworthy to mention that the function $l(r)$ increases monotonically with `r' for different values of $m^2c_1$.

\subsection{Model-I by choosing $\Phi'=0$}
The easiest method is to assume that WHs with zero tidal forces that implies $\Phi'(r) = 0$. The expressions for pressure in this setting, take the following form:
 \begin{eqnarray*}
     % 8\pi\rho &=& \frac{b^{\prime}}{r^{2}}+\frac{m^2C(c_2C+c_1r)}{r^2}   ,    \\
        8\pi p_r &=& -\frac{b}{r^{3}}-\frac{m^2C(c_2C+c_1r)}{r^2} ,  \\
        8\pi p_t  &=& -\left(1-\frac{b}{r}\right)\frac{b^{\prime}r-b}{2r^2(r-b)}-\frac{m^2c_1C}{2r}  ,
  \end{eqnarray*}
In this instance, the expressions of $p_r$ and $p_t$ can be found by using the expressions for $b(r)$ as,
%\begin{widetext}
\begin{eqnarray}
p_r&=&-\frac{1}{
 16 (\beta-3) \pi r^3}\Big[2 (\beta-3) C^2 m^2c_2 r_0 \nonumber\\&&+ (\beta-3) C m^2c_1 (r^2 + r_0^2) -
  16 \pi r^3 \left(\frac{r_0}{r}\right)^{\beta} \rho_c \nonumber\\&&+ 2 r_0 (\beta-3 + 8 \pi r_0^2 \rho_c)\Big],\\
 p_t&=&\frac{1}{32 (\beta-3) \pi r^3}\Big[2 (\beta-3) C^2 m^2c_2 r_0 \nonumber\\&&- (\beta-3) C m^2c_1 (r^2 - r_0^2) - 16 (\beta-2) \pi r^3 \left(\frac{r_0}{r}\right)^{\beta} \rho_c\nonumber\\&& +
 2 r_0 (\beta-3 + 8 \pi r_0^2 \rho_c)\Big].
\end{eqnarray}
%\end{widetext}

\subsection{Model-II by choosing $\Phi=\frac{A}{r}$}
In this subsection, we choose the redshift function of the form:
\begin{eqnarray}
\Phi&=&\frac{A}{r},
\end{eqnarray}
where $A$ is a constant. The redshift function $\Phi(r)$ mentioned above meets the no horizon criterion, permitting two-way travel, since it is both finite and nonzero everywhere. Furthermore, when $r$ approaches to $\infty$, $\Phi(r)$ tends to `zero', making this function asymptotically flat. Kar and Sahdev \cite{Kar:1995vm} used this redshift function to obtain spherically symmetric, static traversable wormholes with traceless matter by solving the tracelessness constraint. Both Morris Thorne-type and Visser-type geometries are discussed with this choice of redshift function.
Assuming this redshift function, in the background of $f(R)$ gravity, Shamir and Fayyaz \cite{Shamir:2020uzy} obtained wormhole geometry by applying the Karmarkar condition for static traversable wormhole geometry. Furthermore, wormhole solutions in several modified theories of gravity, such as $f (R,\,\phi)$ gravity \cite{Malik:2023mte}, and $f(Q)$ gravity \cite{Mustafa:2021ykn}, have been studied with this particular redshift function.\\
Employing the expression of the redshift function, the radial and transverse component of the pressure can be obtained as,
\begin{widetext}
\begin{eqnarray}
p_r&=&-\frac{1}{8 \pi r^3} \bigg[r_0 + C^2  m^2c_2 r_0 + \frac{1}{2} C m^2c_1 (r^2 + r_0^2) + \frac{
    8 \pi \left(r_0^3 - r^3 \left(\frac{r_0}{r}\right)^{\beta}\right) \rho_c}{-3 + \beta} +
    A \Big\{2 + 2 C^2 m^2 c_2 + C m^2 c_1 r \nonumber\\&&- \frac{
       r_0 \big(2 + C m^2 (2 C c_2 + c_1 r_0)\big)}{r} - \frac{
       16 \pi r_0^3 \rho_c}{(-3 + \beta) r} + \frac{
       16 \pi r^2 \left(\frac{r_0}{r}\right)^{\beta} \rho_c}{-3 + \beta}\Big\}\bigg],\\
%\end{eqnarray}
%\end{widetext}
%\begin{widetext}
%\begin{eqnarray}
p_t&=&\frac{1}{32 (
    \beta-3) \pi r^5} \bigg[2 A^2 \Big\{2 (-3 + \beta) (1 + C^2 m^2c_2) r + (\beta-3) C m^2c_1 r^2 - (\beta-3) r_0 \left(2 +
         C m^2 (2 C c_2 + c_1 r_0)\right) \nonumber\\&&- 16 \pi r_0^3 \rho_c +
      16 \pi r^3 \left(\frac{r_0}{r}\right)^{\beta }\rho_c\Big\} +
   A r \Big\{4 (\beta-3) (1 + C^2 m^2c_2) r + (\beta-3) C m^2c_1 r^2 -
      3 (\beta-3) r_0 \times\nonumber\\&&\big(2 + C m^2 (2 C c_2 + c_1 r_0)\big) -
      48 \pi r_0^3 \rho_c + 16 \beta \pi r^3 \left(\frac{r_0}{r}\right)^{\beta} \rho_c\Big\} +
   r^2 \Big\{2 (\beta-3) C^2 m^2c_2 r_0 \nonumber\\&&- (\beta-3) C m^2c_1 (r -
         r_0) (r + r_0) - 16 (\beta-2) \pi r^3 \left(\frac{r_0}{r}\right)^{\beta} \rho_c +
      2 r_0 (\beta-3 + 8 \pi r_0^2 \rho_c)\Big\}\bigg].
\end{eqnarray}
\end{widetext}
\section{Energy Condition}\label{sec4}
\begin{figure*}
    \centering
    \includegraphics[height=6.2cm,width=8.6cm]{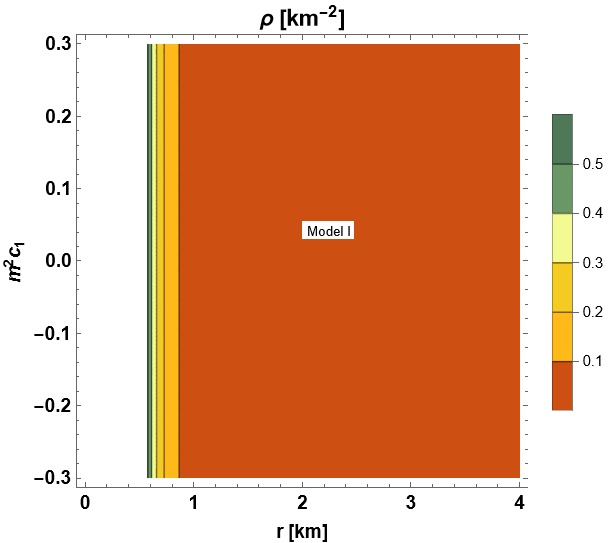}
    \includegraphics[height=6.2cm,width=8.6cm]{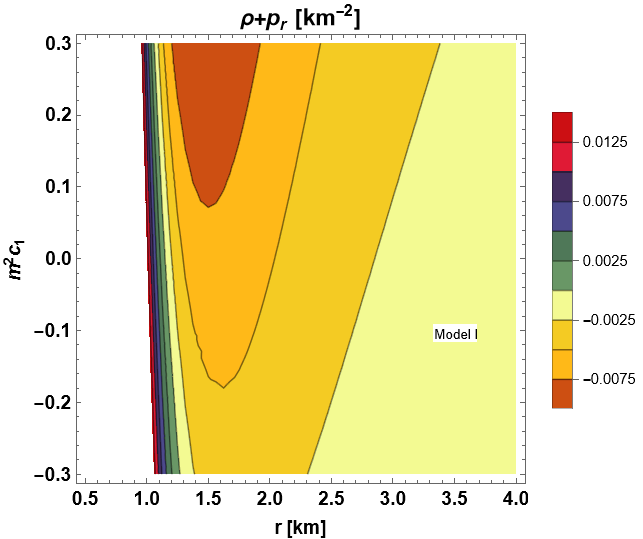}
    \includegraphics[height=6.2cm,width=8.6cm]{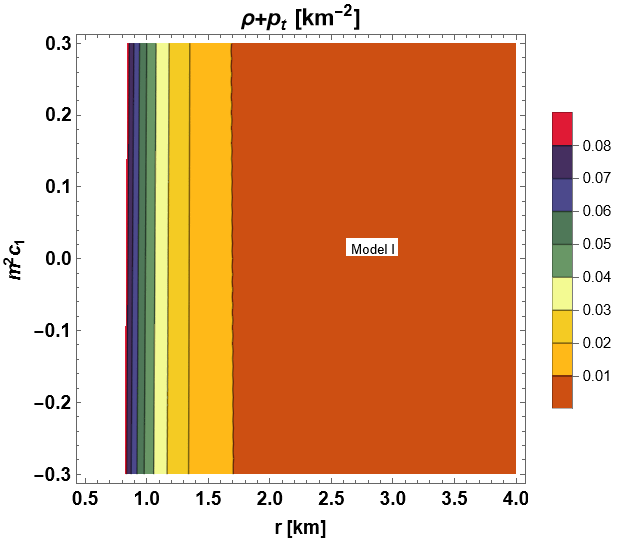}
    \includegraphics[height=6.2cm,width=8.6cm]{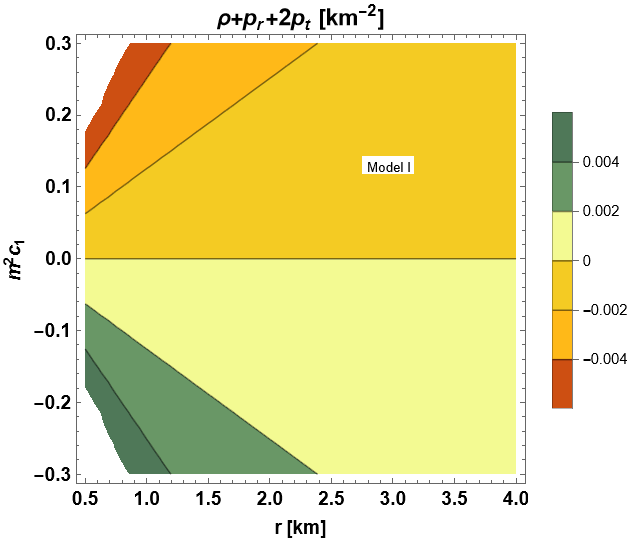}
     \includegraphics[height=6.2cm,width=8.6cm]{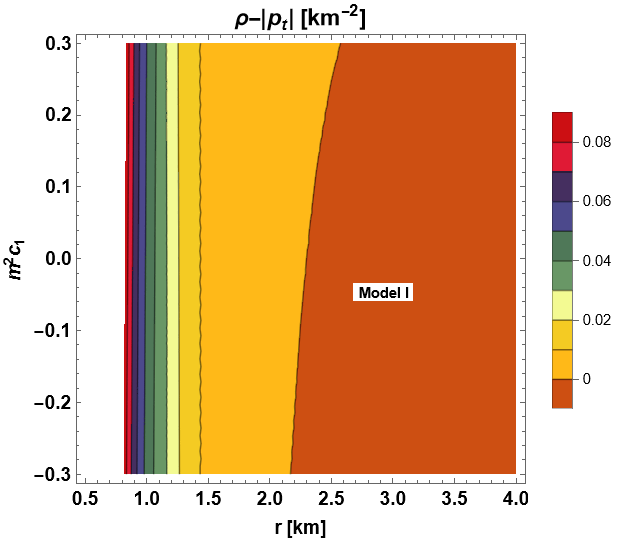}
    \includegraphics[height=6.2cm,width=8.6cm]{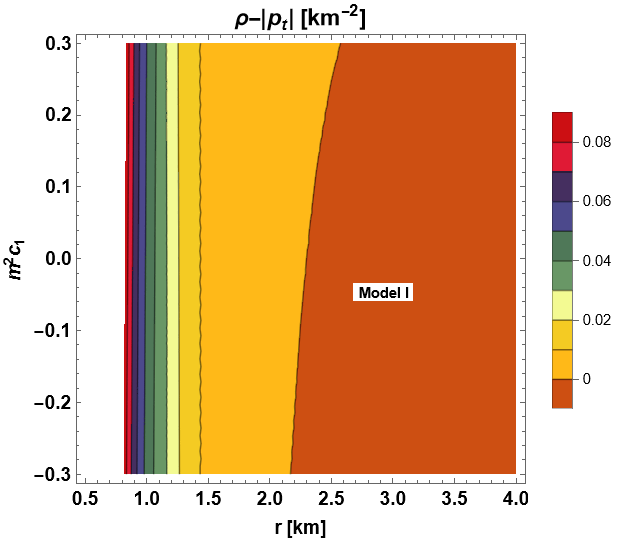}
        \caption{Different types of energy conditions for the model-I are shown against `r'. For drawing the plots we have taken the values of the parameter as $r_0 = 1.5,\, C = 0.4,\,m^2c_2 =0.02,\,\rho_c = 0.011$, and $\beta = 4$.\label{en1}}
\end{figure*}

\begin{figure*}
    \centering
     \includegraphics[height=6.2cm,width=8.6cm]{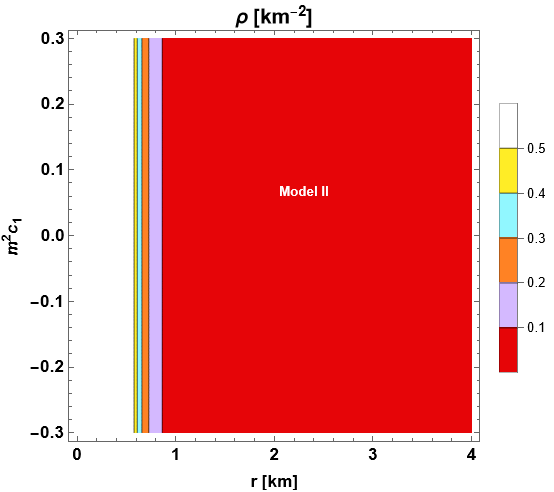}~~~
    \includegraphics[height=6.2cm,width=8.6cm]{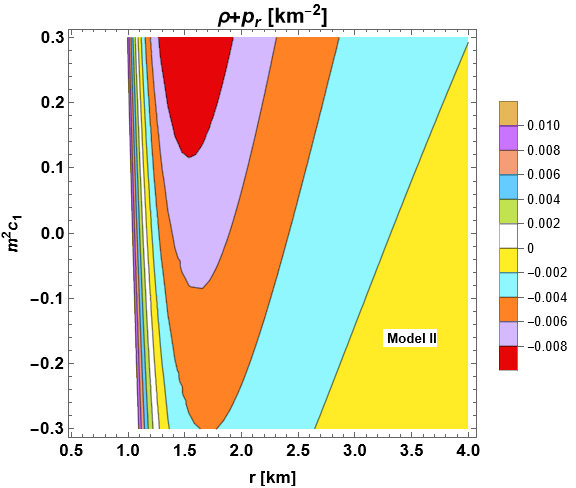}\\
    \includegraphics[height=6.2cm,width=8.6cm]{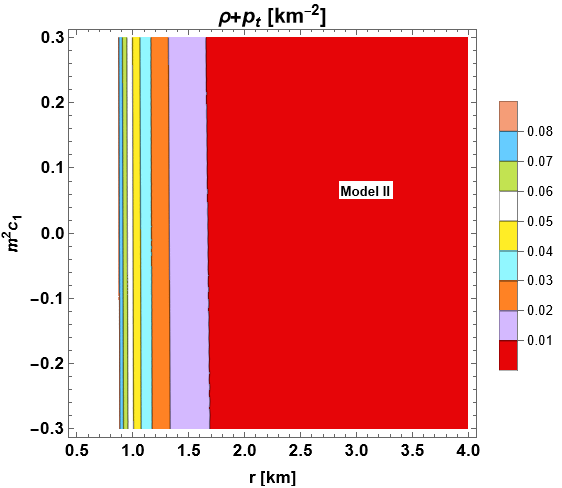}~~~
\includegraphics[height=6.2cm,width=8.6cm]{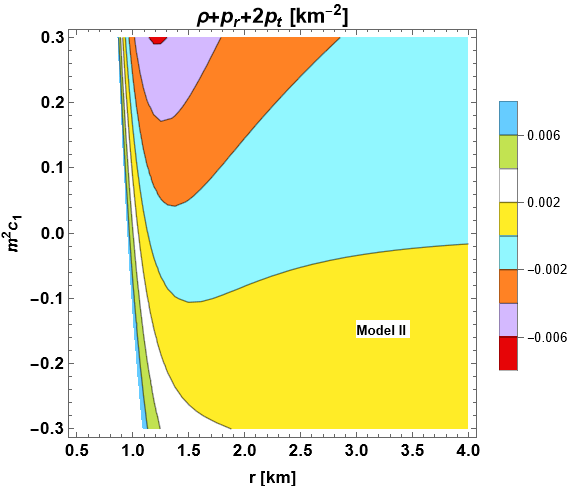}\\
\includegraphics[height=6.2cm,width=8.6cm]{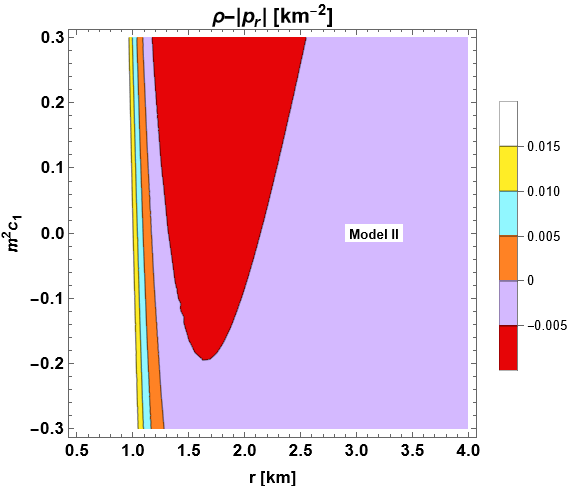}~~~
\includegraphics[height=6.2cm,width=8.6cm]{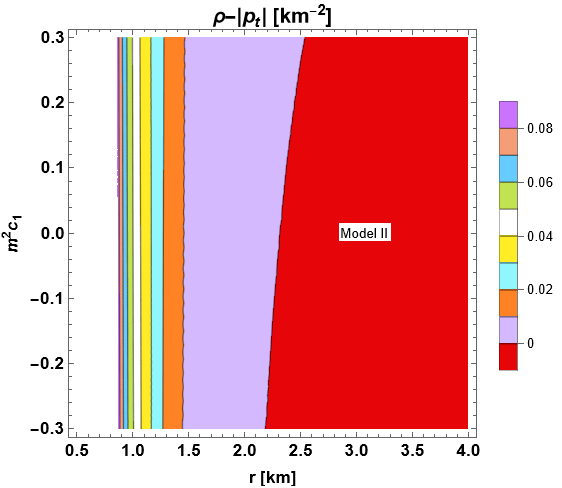}
        \caption{Different types of energy conditions for the model-II are shown against `r'. For drawing the plots we have taken the values of the parameter as $A=0.3,\,r_0 = 1.5,\,C = 0.4,\,m^2c_2 =0.02,\,\rho_c = 0.011$, and $\beta = 4$.\label{en2}}
\end{figure*}

Energy Conditions are essential to comprehending wormhole structure since their violation may suggest the presence of exotic matter. Energy conditions can be broadly classified into four categories: null energy condition (NEC), weak energy condition (WEC), strong energy condition (SEC), and dominant energy condition (DEC).
The energy conditions in connection with matter density and pressures are stated as \cite{Morris:1988cz,Morris:1988tu,Visser:1995cc},
\begin{itemize}
    \item According to the strong energy condition (SEC), gravity must always be attractive. It is defined by $(T_{\mu\nu}-\frac{T}{2}g_{\mu\nu})U^{\mu}U^{\nu}\geq 0$, where $T$ is the trace of the stress-energy tensor, $U^{\nu}$ is the time-like vector. In terms of principal pressures it can be expressed as $\rho+p_r\geq0$, $\rho+p_t\geq0$, $\rho+p_r+2p_t\geq0$,
   \item  According to the dominant energy condition (DEC), any observer's measurement of the energy density should be non-negative, or $T_{\mu\nu}U^{\mu}U^{\nu}\geq 0$, $T_{\mu\nu}U^{\mu}$ is not space-like, which results in $\rho \geq0$, $\rho- |p_r|\geq 0$, $\rho- |p_t| \geq 0$,
    \item Any observer's measurement of the energy density must always be non-negative, according to the weak energy condition (WEC), which implies
 $T_{\mu\nu}U^{\mu}U^{\nu}\geq 0$, $U^{\mu}$ being timelike vector. In terms of principal pressures $\rho \geq 0$, $\rho+p_r\geq0$, $\rho+p_t\geq0$,
\item One basic criterion from SEC and WEC is the null energy condition (NEC) that leads to, $T_{\mu\nu}k^{\mu}k^{\nu}\geq 0$, $k^{\mu}$ being null-like vector, in terms of principal pressures, $\rho+p_r \geq 0$, $\rho+p_t \geq 0$.
\end{itemize}
To verify the conditions mentioned above, we produced profiles for both models, and the figures are displayed in Figs.~\ref{en1}-\ref{en2}.\\
\begin{itemize}
  \item It can be seen in Figs.~\ref{en1}-\ref{en2}, the energy density is positive,i.e., $\rho \geq 0$, for both the models when $m^2c_1\in [-0.3,\,0.3]$ and for all $r>0$.
  \item In Figs.~\ref{en1}-\ref{en2}, we see that $\rho+p_r>0$ for $r \in [0,\,1.5]$ and $\rho+p_r<0$ when $r>1.5$, and when $m^2c_1\in [-0.3,\,0.3]$.
 \item All of the quantities $\rho+p_t,\,\rho-|p_r|$, and $\rho-|p_t|$ all are positive for all $r>0$ and when $m^2c_1\in [-0.3,\,0.3]$ as can be seen in Figs.~\ref{en1}-\ref{en2}.
\item We can see that $\rho+p_r+2p_t<0$ for all $r>0$ and when $m^2c_1\in [-0.3,\,0.3]$ in Figs.~\ref{en1}-\ref{en2}.
\end{itemize}
In a nutshell, for $m^2c_1\in [-0.3,\,0.3]$ and $r>1.5$, except DEC, all three energy conditions namely NEC, WEC, and SEC are violated. This demonstrates the presence of exotic matter in wormhole spacetime in the background of massive gravity for both models.

\section{Deflection angle of photons}\label{sec5}
\begin{figure}[htbp]
    \centering
     \includegraphics[scale=.5]{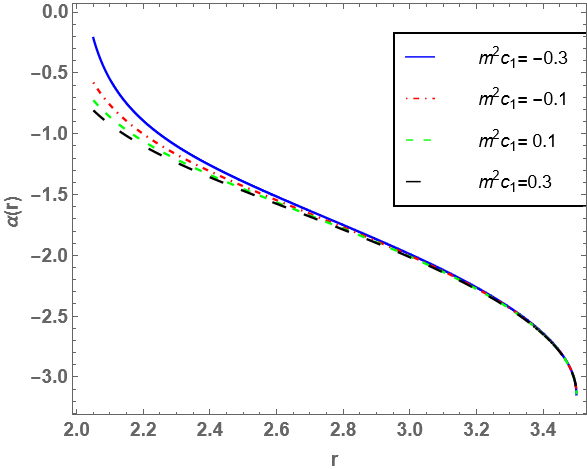}
        \caption{The Deflection angle is shown against `r'. For drawing the plots we have taken the values of the parameter as $A=0.3,\,r_0 = 1.5,\, C = 0.4,\,m^2c_2 =0.02,\,\rho_c = 0.011$, and $\beta = 4$.\label{alphar}}
\end{figure}

To verify the deflection angle of photons from null geodesics, we first introduce a spherically symmetric and static line element, provided by,
\begin{equation}\label{eq38}
ds^2 = - A(r) dt^2 + B(r) dr^2 + C(r) d\Omega^2.
\end{equation}
The momenta one-forms of a freely falling body and the background geometry are related by the geodesic equation, which is given by \cite{schutz},
\begin{equation}\label{eq39}
\frac{dp_\beta}{d\lambda} = \frac{1}{2} g_{\nu \alpha, \beta}~p^\nu p^\alpha,
\end{equation}
$\lambda$ being the affine parameter. $p_\beta$ is a constant of motion if for a fixed index $\beta$, $g_{\alpha\nu}$ are independent of $x^\beta$. Therefore, in Eq. (\ref{eq39}), all of the $g_{\alpha\beta}$ are independent of $t,\,\theta$, and $\phi$ if we just take into account the equatorial slice by setting $\theta=\frac{\pi}{2}$. Using $\alpha$ as a cyclic coordinate, one can obtain the corresponding killing vector fields $ \delta^{\mu}_{\alpha} \partial_\nu $. The constants of motion $p_t$ and $p_\phi$ can now be adjusted to
\begin{equation}\label{eq40}
	p_t = -E, ~~~~~~~~~~~~~ p_\phi = L,
\end{equation}
respectively. Where $ E $ and $ L $ denote the photon's energy and angular momentum, respectively.
Therefore, we have
\begin{eqnarray}\label{eq41}
\nonumber	p_t = \dot{t} = g^{t \nu} p_\nu = \frac{E}{A(r)}, \\
	p_\phi = \dot{\phi} = g^{\phi \nu} p_\nu = \frac{L}{C(r)},
\end{eqnarray}
where the differentiation concerning the affine parameter $\lambda$ is represented by the overdot. Furthermore, one can obtain the expression for radial null geodesic as,
\begin{equation}\label{eq42}
	\dot{r}^2 = \frac{1}{B(r)} \left( \frac{E^2}{A(r)} - \frac{L^2}{C(r)} \right).
\end{equation}
In terms of the impact parameter $\mu=\frac{L}{E}$, the equation for the photon trajectory can be expressed as,
\begin{equation}\label{eq43}
	\left( \frac{dr}{d\phi} \right)^2 = \frac{C(r)^2}{\mu^2 B(r)} \left[ \frac{1}{A(r)} - \frac{\mu^2}{C(r)} \right].
\end{equation}
The deflection angle of a photon may now be obtained by considering the source of photon radius $r_s$ that is producing the geometry. Photons can only reach the surface when an existing solution $r_0$ satisfies the requirements $r_0 > r_s$ and $\dot{r}^2=0$.
In this case, $r_0$ represents the turning point or distance of the closest approach. In such a scenario, the impact parameter therefore becomes
\begin{equation}\label{eq44}
	\mu = \frac{L}{E} = \pm \sqrt{\frac{C(r_0)}{A(r_0)}},
\end{equation}
and in weak gravity limit, $\mu\approx\sqrt{C(r_0)}$. According to Bhattacharya and Potapov \cite{Bhattacharya:2010zzb}, the deflection angle of a photon coming from the polar coordinate limit $ \lim\limits_{r \rightarrow \infty} \left( r, -\frac{\pi}{2}-\frac{\alpha}{2} \right) $, after passing through the turning point at $ (r_0, 0) $ and approaching $ \lim\limits_{r \rightarrow \infty} \left( r, \frac{\pi}{2}+\frac{\alpha}{2} \right)$, is this $ \alpha $, which is a function of $ r_0 $. Using Eq.~\eqref{eq43}, we can calculate it in this way:
\begin{equation}\label{eq45}
	\alpha(r_0) = -\pi + 2 \int_{r_0}^{\infty}
	\frac{\sqrt{B(r)} dr}{\sqrt{C(r)} \left[ \left( \frac{A(r_0)}{A(r)} \right) \left( \frac{C(r)}{C(r_0)} \right) -1 \right]^{1/2}},
\end{equation}
For the selection of metric coefficients in wormhole geometry, the deflection angle becomes,
\begin{equation}\label{eq46}
	\alpha(r_0)=-\pi+2 \int_{r_0}^{\infty} \frac{dr}{r \left[ \left( 1- \frac{b(r)}{r} \right) \left( \frac{r^2}{r_0^2} -1 \right) \right]^{1/2} }.
\end{equation}
Now, by numerically integrating the Eq.~(\ref{eq46}) after imposing the shape function provided in Eq.~(\ref{br}), we have displayed the deflection angle of photons in massive gravity which is shown in Fig.~\ref{alphar}. Notably, we find that at a particular value of $r_0$, the deflection angle becomes negative. This can be regarded as one of the special characteristics of the dRGT massive gravity, which is the repulsive effect caused by gravity. Similar type of results have been previously obtained by Panpanich et al. \cite{Panpanich:2019mll} when describing the particle motions and gravitational lensing in dRGT massive gravity theory, Bhar et al. \cite{Bhar:2024vov} when describing the model of non-commutative geometry in dRGT massive gravity theory, Dutta et al. \cite{Dutta:2023wfg} when analyzed the wormhole solution in the dRGT-f(R,\,T) massive gravity. Note that modified gravity theories involving exotic matter and energy \cite{Kitamura:2012zy,Izumi:2013tya,Kitamura:2013tya,Shaikh:2017zfl} and BHT massive gravity \cite{Nakashi:2019jjj} also exhibit negative deflection angles.

\section{Measurement of exotic matter}\label{sec6}
The ``volume integral quantifier" (VIQ), as described by Visser et al. \cite{Visser:2003yf}, and Kar et al. \cite{Kar:2004hc} can be used to quantify the total amount of matter that violates the null energy condition. In other words, the amount of `exotic matter' can be measured by the magnitude of the negative value of VIQ. Additionally, it has been demonstrated in Refs.~\cite{Visser:2003yf,Kar:2004hc} that quantum physics is known to lead to small violations of the averaged NEC while also forming traversable WHs with arbitrarily small amounts of averaged NEC-violating matter. The term exotic matter is typically used to describe material that violates the NEC. The following VIQ can be used to compute the amount of exotic matter near the throat of the wormhole:
\begin{eqnarray}\label{viq12}
I_V=\oint(\rho+p_r)dV,
\end{eqnarray}
where the volume can be written as $dV=r^2dr d\Omega$, $d\Omega$ being the solid angle. Since $\oint dV=2\int_{r_0}^{\infty}dV=8\pi\int_{r_0}^{\infty}r^2dr$, we can write eqn. (\ref{viq12}) as,
\begin{eqnarray}
I_V=8\pi\int_{r_0}^{\infty}(\rho+p_r)r^2dr.
\end{eqnarray}
In terms of the wormhole metric coefficients, the above integral (\ref{viq12}) can be written as,
\begin{eqnarray}
I_V&=&-\int_{r_0}^{\infty}(1-b')\ln\left[e^{2\Phi}\left(1-\frac{b(r)}{r}\right)^{-1}\right]dr,\nonumber\\
\end{eqnarray}
For our present models, with a cut-off of the stress-energy at $a_1>r_0$, the above integral leads to,
\begin{widetext}
\begin{eqnarray}
I_V\left|\right._{\text{Model-I}}&=&-\frac{1}{4 ( \beta-3)^2} \Big[r^2 \Big\{(\beta-3)^2 C m^2c_1 +
    32 (\beta-2) \pi r  \left(\frac{r_0}{r}\right)^{\beta}\rho_c\Big\} +
 2 (\beta-3) r_0 \Big\{(\beta-3) \big(2 + C m^2 (2 C c_2 + c_1 r_0)\big)\nonumber\\&& +
    16 \pi r_0^2 \rho_c\Big\} \log[r]\Big]_{r_0}^{a_1},\label{v1}\\
%\end{eqnarray}
%\end{widetext}
%\begin{widetext}
%\begin{eqnarray}
I_V\left|\right._{\text{Model-II}}&=&\bigg[- \frac{1}{4}Cm^2c_1 r^2 + \frac{16 A \pi r^2 \left(\frac{r_0}{r}\right)^{\beta} \rho_c}{
 6 - 5 \beta + \beta^2} - \frac{
 8 (-2 + \beta) \pi r^3 \left(\frac{r_0}{r}\right)^{\beta} \rho_c}{(-3 + \beta)^2} - \frac{
 A r_0 \big\{2 + C m^2 (2 C c_2 + c_1 r_0) + \frac{
    16 \pi r_0^2 \rho_c}{-3 + \beta}\big\}}{r}\nonumber\\&& - 2 A (1 + C^2 m^2c_2) \log[r] -
 \frac{1}{2} r_0 \big\{2 + C m^2 (2 C c_2 + c_1 r_0)\big\} \log[r] - \frac{
 8 \pi r_0^3 \rho_c \log[r]}{-3 + \beta}-A C m^2c_1 r\bigg]_{r_0}^{a_1}.\label{v2}
\end{eqnarray}
\end{widetext}
\begin{figure*}
    \centering
     \includegraphics[height=6.5cm,width=7.5cm]{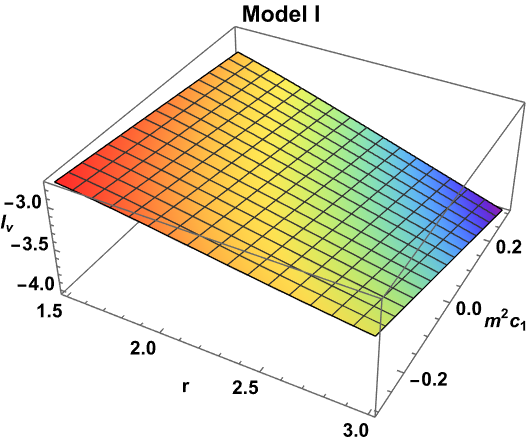}
   \includegraphics[height=6.5cm,width=7.5cm]{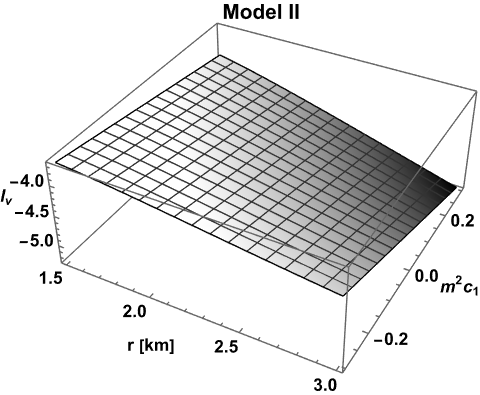}
        \caption{The volume integral quantifier is shown against `r'. For drawing the plots we have taken the values of the parameter as $A=0.3,\,r_0 = 1.5,\, C = 0.4,\,m^2c_2 =0.02,\,\rho_c = 0.011$, and $\beta = 4$.\label{viq1}}
\end{figure*}
using expression (\ref{v1})-(\ref{v2}), the fluctuation of VIQ concerning `r' is shown in Fig.~\ref{viq1}.
Eqns.~(\ref{v1})-(\ref{v2}) allow us to determine that, when taking the limit $a_1\rightarrow r_0$, $I_V\rightarrow0$. The figure shows that, in the given range of $-0.3<m^2c_1<0.3$, it is theoretically possible to construct a traversable WH with a small quantity of NEC-violating matter.
From the figures, it can be seen that $I_V<0$, suggests the existence of exotic materials at the throat of the WH.
\section{Complexity factor}\label{sec7}
According to Herrera \cite{Herrera:2018bww}, in the case of a spherically symmetric spacetime, the complexity factor is defined by,
\begin{eqnarray}\label{c1}
Y_{TF}&=&8\pi(p_r-p_t)-\frac{4\pi}{r^3}\int_{0}^{r}x^3\rho'(x)dx,
\end{eqnarray}
According to the formula, $Y_{TF}=0$ for an isotropic and homogeneous matter distribution.
It can be established that the Eqn. (\ref{c1}) helps us to write the Tolman mass as follows:
\begin{eqnarray}
m_T&=&(m_T)_{\Sigma}\left(\frac{r}{r_{\Sigma}}\right)^3+r^3\int_0^{r_{\Sigma}}\frac{e^{\frac{\Phi+\lambda}{2}}}{x}Y_{TF}dx,
\end{eqnarray}
\begin{figure*}[htbp]
    \centering
     \includegraphics[height=6.5cm,width=8cm]{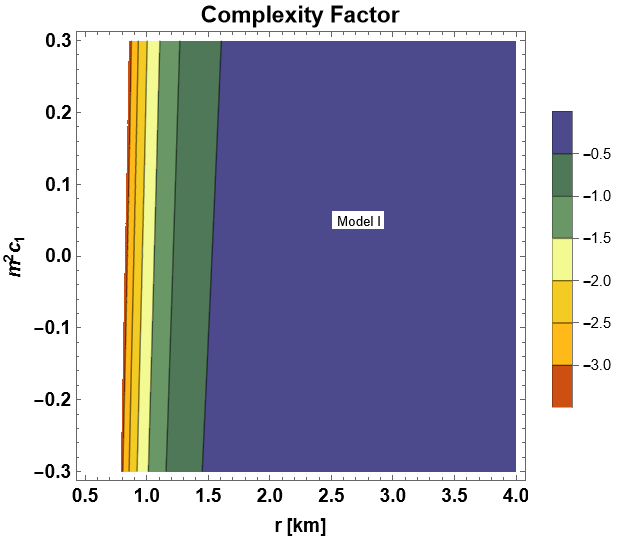}~~~
   \includegraphics[height=6.5cm,width=8cm]{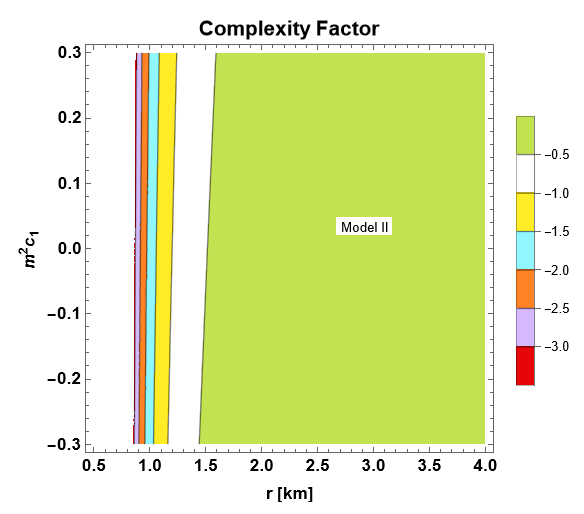}
        \caption{The complexity factor is shown against `r'. For drawing the plots, we have taken the values of the parameter as $A=0.3,\,r_0 = 1.5,\, C = 0.4,\,m^2c_2 =0.02,\,\rho_c = 0.011$, and $\beta = 4$.\label{complex}}
\end{figure*}
It should be noted that the vanishing complexity criterion ($Y_{TF}=0$) not only holds in the case of isotropic and homogeneous systems, but it also holds in all situations where,
\begin{eqnarray}\label{com2}
p_r-p_t&=&\frac{1}{2r^3}\int_{0}^{r}x^3\rho'(x)dx,
\end{eqnarray}
By definition, the only thing that can support wormholes is a non-homogeneous distribution of matter. Therefore, the situation that arises in Eqn.~(\ref{com2}) is the only one where a zero complexity wormhole can be achieved. The vanishing complexity criterion yields a non-local equation of state that may be applied as a supplementary requirement to close the system of EFEs \cite{Arias:2022qrm,Casadio:2019usg,Contreras:2021xkf,Avalos:2022tqg}. We now plot the diagrams for each previously obtained WH to determine the range of $Y_{TF}$. The nature of the complexity factor for every WH is displayed in Fig.~\ref{complex}. It is evident that the range of $Y_{TF}$ depends on the values of $m^2c_1$ and is provided by $-3.5<Y_{TF}<0$. We can see that, the complexity factor is increasing monotonically in the vicinity of the wormhole throat, and $Y_{TF}$ is getting closer to zero for higher values of the radial coordinate. It can be concluded that as $r\rightarrow \infty$, the effect of anisotropic pressure and inhomogeneous energy density cancel one another which results in the zero complexity factor.
\section{TOV equation}\label{sec8}
The stability of the wormhole geometry can be established by using the equilibrium condition and the generalized Tolman-Oppenheimer-Volkov (TOV) equation \cite{Tolman:1939jz,Oppenheimer:1939ne}. This equation is obtained from the conservation law of energy and it is given as follows:
\begin{eqnarray}\label{tov1}
-\frac{dp_r}{dr}-\frac{d\Phi}{dr}(\rho+p_r)+\frac{2}{r}(p_t-p_r)=0,
\end{eqnarray}
\begin{figure*}[htbp]
    \centering
     \includegraphics[height=6.6cm,width=8.6cm]{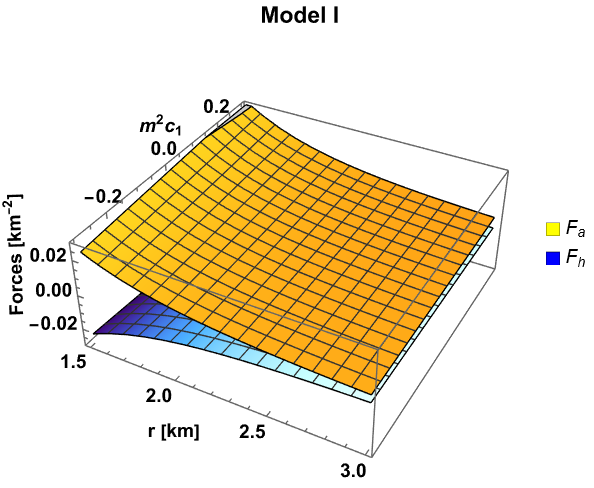}
   \includegraphics[height=6.6cm,width=8.6cm]{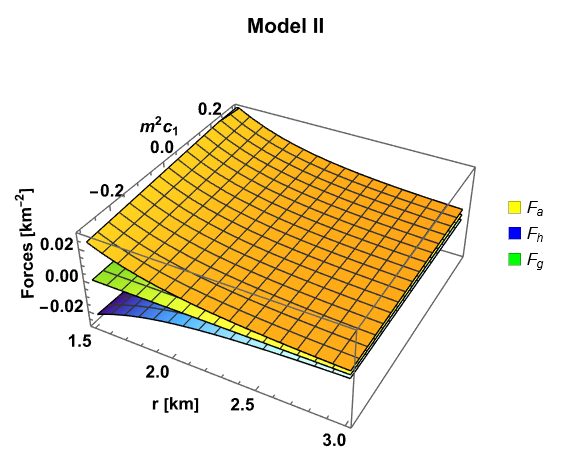}
        \caption{Different forces acting on the wormhole model. For drawing the plots we have taken the values of the parameter as $A=0.3,\,r_0 = 1.5,\, C = 0.4,\,m^2c_2 =0.02,\,\rho_c = 0.011$, and $\beta = 4$.\label{tov3}}
\end{figure*}
From the standpoint of equilibrium, we shall now try to understand the previous equation (\ref{tov1}) by observing it as a combination of three distinct forces: gravitational ($F_g$), hydrostatic ($F_h$), and anisotropic ($F_a$) forces. The expressions of the forces are listed below:
\begin{eqnarray}
F_g&=&-\frac{d\Phi}{dr}(\rho+p_r) ,\\
F_h&=&-\frac{dp_r}{dr} ,\\
F_a&=&\frac{2}{r}(p_t-p_r),
\end{eqnarray}
and for a static equilibrium, we need the sum of the three forces to be equal to zero. A graphical representation of all these different forces for a wide range of $m^2c_1$ is presented in Fig.~\ref{tov3}. For a wormhole to remain stable, the sum of the forces mentioned earlier must equal zero. Fig.~\ref{tov3} clearly illustrates this. Therefore, for model I, the anisotropic and hydrostatic forces balance each other. On the other hand, model II yields a stable, perfectly balanced wormhole solution because the combined action of hydrostatic and gravitational forces balances the anisotropic forces and stabilizes the forces surrounding the wormhole.
\section{Discussion}\label{sec9}
Using an anisotropic energy-momentum tensor (EMT) and a particular form of matter density function, in the present work, we developed models of static wormholes within the framework of dRGT massive gravity. By assuming both constant and variable redshift functions, we were able to derive two distinct spherically symmetric wormhole solutions in this modified gravity. Through the utilization of Eqn.~(\ref{rho}), we were able to derive the shape function of the wormhole as well as the material solutions for the WH, specifically the transverse pressure $p_t$ and the radial pressure $p_r$ for each of the two models. Within the context of the considered gravitational theory, these equations give the formulas for the physical quantities characterizing the matter composition of the WH. The four basic energy conditions (ECs) NEC, WEC, DEC, and SEC were reviewed in the current article. The presence of non-exotic matter within the WH depends on the respectability of the ECs, especially the NEC, WEC, DEC, and SEC throughout the entire WH spacetime. This is a desired property since it guarantees that the matter content that fills the WH does not break ECs or have a negative energy density and is consistent with ordinary matter. Evidently, for all values of r, the NEC is violated. The violation of NEC indicates that exotic matters may be present in the wormhole spacetime with unusual properties. This exotic matter may not meet the ECs required by normal matter, or it might have a negative energy density. Since WH solutions commonly require such exotic matter to preserve the WH's structure, violation of NEC is a typical feature of the wormhole. Researchers are still actively examining the existence and characteristics of such unusual matters. According to our research, a negative deflection angle is produced as a result of repulsive gravity. We introduced the idea of the photon deflection angle on the WH to validate this phenomenon. This deflection angle in spacetime becomes negative when photons are subjected to repulsive gravity, as reported in \cite{Panpanich:2019mll}. In addition, it is important to remember that for all values of $m^2c_1$, the deflection angle continuously maintains negative values. This continuous negative value of the deflection angle may be considered as an example of the repulsive gravity effect. Therefore, the general characteristic of dRGT massive gravity and the source of massive gravitons are the non-asymptotic flatness and the presence of repulsive gravity.
It has been demonstrated that, in theory, traversable wormhole solutions can be developed with an arbitrarily small amount of energy condition-violating matter. We have also taken into consideration the ``volume integral quantifier", which offers helpful information regarding the total amount of energy condition-violating matter. We can therefore conclude that a stable traversable wormhole solution can be obtained by effectively minimizing the use of exotic matter through modification of standard GR. With the use of graphical representation, we have also described the complexity factor of our current model. As seen in Fig.~\ref{complex}, the complexity factor for Wormhole Models I and II is monotonically increasing in the vicinity of the wormhole throat, and $Y_{TF}$ is approaching zero for higher values of the radial coordinate. We took into consideration the Tolman-Oppenheimer-Volkov (TOV) equation from the perspective of the equilibrium condition for wormhole configuration. Three components contribute to the equation: the hydrostatic force $F_h = -\frac{dp_r}{dr}$, the anisotropic force $F_a=\frac{2}{r}(p_t-p_r)$, and the gravitational force $F_g=-(\rho+ p_r)\frac{d\Phi}{dr}$. The last component disappears for model I since the redshift function is taken to be constant. Since all of the forces add up to zero, we may infer that the wormhole solutions are in equilibrium with the aid of the graphical representation.\par
The existence of exotic matter is the main challenge to the physical viability of a traversable wormhole. Eventually, our goal should be to find wormhole structures that do not require exotic matter to sustain the model. For the obtained solutions, intriguing features like gravitational lensing and associated shadow casting can also be studied in the future. The presence of photon spheres and the existence of charge can also be studied under different conditions. This work offers a more comprehensive generalization to Morris-Thorne wormholes in General Relativity, which are included as a special case in the limits $m \rightarrow 0$.
\section*{Acknowledgement}

P.B. is thankful to the Inter-University Centre for Astronomy and Astrophysics (IUCAA), Pune, Government of India, for providing visiting associateship.

\end{document}